\documentclass[aps,prd,onecolumn,nofootinbib,superscriptaddress,floatfix]{revtex4-2}

\usepackage{amsmath,amssymb,mathtools}
\usepackage{bm}
\usepackage{graphicx}
\usepackage{siunitx}
\usepackage{booktabs}
\usepackage{xcolor}
\usepackage[colorlinks=true,linkcolor=blue!50!black,citecolor=blue!50!black,urlcolor=blue!50!black]{hyperref}

\newcommand{\R}{\mathbb{R}}

\newcommand{\chimin}{\chi_{\min}}

\newcommand{\lo}{\mathrm{lo}}
\newcommand{\hi}{\mathrm{hi}}
\newcommand{\half}{h}

\newcommand{\sstat}{\sigma_{\rm stat}}
\newcommand{\ssys}{\sigma_{\rm sys}}
\newcommand{\stot}{\sigma_{\rm tot}}

\newcommand{\code}[1]{\texttt{#1}}

\DeclareMathOperator{\Prob}{Prob}



\begin{document}

\title{Interpretable uncertainties for bounds on spectral functionals\\
from Euclidean correlators}

\author{Christoph Lehner}
\affiliation{Fakult\"at f\"ur Physik, Universit\"at Regensburg, Universit\"atsstra{\ss}e 31, 93040 Regensburg, Germany}
\email{christoph.lehner@ur.de}

\date{\today}

\begin{abstract}
Positivity of the spectral density, combined with a $\chi^2$ measure of compatibility with the
data, bounds every linear functional of the spectral density of a Euclidean correlator to an
interval.  We show how such an interval is turned into a central value with statistical and
systematic errors that have a definite meaning.  We first show that, at typical precision, the best positive
fit of noisy data has $\chi^2_{\min}>0$, with a probability bounded from below by the distance
of the noise-free data from the boundary of the set of data of positive spectra.  Numerically,
it does not depend on the number of states.  The threshold must therefore be set relative to $\chi^2_{\min}$.
With $\Delta\chi^2=1$, the interval reproduces Gaussian error propagation when positivity
does not act, and positivity can only narrow it.  We define the statistical error by the
jackknife and the systematic error as the remaining, non-Gaussian part of the interval, which
becomes a systematic error that does not decrease with the statistics in the high-statistics
limit.  Coverage tests for correlators and for smeared spectral densities and step functions
give nominal coverage for outputs determined by the data and mostly conservative coverage
otherwise.  The exceptions are effects of the statistical error on the error estimate that
disappear as the statistics is increased.
Additional bounds on the spectral density, for example from perturbation theory, further reduce the
systematic error substantially without loss of coverage.  For long-distance correlators
the method uses the same information as the bounding method, and its error agrees with that
of the bounding method up to the remaining gap between the bounds.
\end{abstract}

\maketitle

\section{Introduction}\label{sec:intro}

Lattice QCD determines Euclidean correlators $C(t)$ precisely only on a window of times: at
short times by cut-off effects, at long times by a signal-to-noise ratio that decreases
exponentially~\cite{Parisi:1984,Lepage:1989}.  Many quantities of interest are linear functionals
of the spectral density $\rho$ that depend on $C(t)$ outside this window, or on $\rho$ itself.
Examples are the correlator at long distances, which enters the hadronic vacuum polarization
contribution to the muon anomalous magnetic moment in the time-momentum
representation~\cite{BerneckerMeyer:2011,RBC:2018}, and spectral densities or their integrals
smeared over an energy range~\cite{Hansen:2019}.  Reconstructing them from the data is an
ill-posed inverse problem.  The linear method of Ref.~\cite{Hansen:2019} computes smeared
spectral densities with a statistical error at fixed smearing width and recovers the target
in the limit of vanishing width.

A complementary approach makes no assumption beyond the properties of the spectral density
itself.  Since $\rho\ge0$, the data restrict every linear functional of $\rho$ to an interval,
which can be computed as a convex optimization problem with certificates from convex
duality~\cite{Lawrence:2024}.  Ref.~\cite{Abbott:2026} develops this approach for smeared
spectral functions, including matrix-valued correlators, with a reduction to semidefinite
programs, and Ref.~\cite{Mutzel:2026} constructs bounds that hold for all spectra rather than
on a grid.  Analyticity leads to related bounds through Nevanlinna--Pick
interpolation~\cite{Bergamaschi:2023,Fields:2025}.  For the long-distance correlator, the
bounding method~\cite{Lehner:2016talk,RBC:2018,BMW:2017slope} is a simple instance of the
same idea: it bounds $C(t)$ beyond a time $T$ from above and below using positivity and the
lowest energy in the spectrum.

Noisy data enter these approaches through a $\chi^2$ measure of compatibility, and the
result is an interval whose width combines the statistical error of the data with the
ambiguity of the spectrum that the data leave open.  Refs.~\cite{Lawrence:2024,Abbott:2026,Mutzel:2026}
admit all spectra with $\chi^2$ below a fixed threshold. Ref.~\cite{Fields:2025} samples data
compatible with both the errors and the constraints and combines the fluctuation of the
interval with its width.  To use such results in precision calculations, combined with other
quantities and their correlations, one needs a central value with a statistical and a
systematic error that have a definite meaning.  The statistical error should be Gaussian to
the extent that the data are, at least where the constraints have little impact, and the total
error should cover the true value with at least the nominal probability where they do.

In this paper we construct errors with these properties.\footnote{Related questions were studied independently in
Ref.~\cite{Abbott:2026cones}, which analyses the conic geometry of
Euclidean data compatible with positivity and the projection of noisy
data onto this set.  The two works were developed in parallel and are
submitted simultaneously.}
After the setup in Sec.~\ref{sec:setup},
we show in Sec.~\ref{sec:chi2min} that at typical precision no positive spectrum reproduces
noisy data exactly, so that the threshold has to be set relative to the best positive fit.  In
Sec.~\ref{sec:stat} we show that the threshold $\Delta\chi^2=1$ above the best fit, which
defines a profile-likelihood interval~\cite{JamesRoos:1975,James:2006}, reproduces Gaussian error
propagation when positivity does not act, and that positivity can only narrow it.  In
Sec.~\ref{sec:split} we split the error into a statistical part, from the jackknife, and a
non-Gaussian remainder, which becomes a systematic error at high statistics, and in
Sec.~\ref{sec:coverage} we test the coverage of the total error.  Section~\ref{sec:prior}
extends the construction to two-sided bounds on the spectral density and relates it to the
bounding method, and Sec.~\ref{sec:conclusions} concludes.

\section{Setup}\label{sec:setup}

We first define the data, the representation of the spectral density, and the smallest
$\chi^2$ of a positive spectrum, on which everything below is built.  In a finite volume the Euclidean correlator $C(t)=\langle O(t)\,O^\dagger(0)\rangle$ has
the spectral representation
\begin{equation}\label{eq:laplace}
  C(t) \;=\; \int_0^\infty d\omega\;\rho(\omega)\,e^{-t\omega},
  \qquad \rho(\omega)\;=\;\sum_n\big|\langle n|O^\dagger|0\rangle\big|^2\,\delta(\omega-E_n)\;\ge\;0 ,
\end{equation}
where the sum runs over the states $|n\rangle$ with energies $E_n$.

The data are estimates $\bar C\in\R^m$ of $C$ at $m$ Euclidean times $t_1<\dots<t_m$,
with covariance $\Sigma$.  We model them as
\begin{equation}\label{eq:model}
  \bar C\;\sim\;\mathcal N(c,\Sigma),\qquad c_i=C(t_i).
\end{equation}

Spectral densities are restricted to an interval $[\omega_1,\omega_K]$ and represented
on a grid $\omega_1<\dots<\omega_K$ by weights $z_j\ge0$,
$\rho(\omega)=\sum_jz_j\,\delta(\omega-\omega_j)$, so that the data of a spectrum are $Ez$
with $E_{ij}=e^{-t_i\omega_j}$.  We assume that the true spectrum is supported in
$[\omega_1,\omega_K]$.  With
\begin{equation}
  \chi^2(z)=(Ez-\bar C)^T\Sigma^{-1}(Ez-\bar C),
\end{equation}
the smallest $\chi^2$ of any positive spectrum is
\begin{equation}\label{eq:chimin}
  \chimin^2\;=\;\min_{z\ge0}\,\chi^2(z),
\end{equation}
a non-negative least-squares problem~\cite{LawsonHanson}.  This combination of
positivity with a $\chi^2$ measure of compatibility with the data is the framework of
Refs.~\cite{Lawrence:2024,Abbott:2026,Mutzel:2026}.  $\chimin>0$ means that no positive
spectrum on the grid reproduces the data exactly.

\section{Interpretability of uncertainties}\label{sec:interp}

In this section we construct a central value with statistical and systematic errors from
positivity and the data.  We first show that at typical precision the best positive fit
does not reproduce noisy data (Sec.~\ref{sec:chi2min}), which determines how the threshold on $\chi^2$
has to be set.  We then show that with this threshold the errors are Gaussian where
positivity does not act (Sec.~\ref{sec:stat}), separate the statistical and systematic parts
in general (Sec.~\ref{sec:split}), and test the coverage of the result (Sec.~\ref{sec:coverage}).

\subsection{Distribution of the minimal $\chi^2$}\label{sec:chi2min}

The data of all positive spectra on $[\omega_1,\omega_K]$ form the convex cone
\begin{equation}
  \mathcal P\;=\;\Big\{\int_{\omega_1}^{\omega_K}d\omega\,\rho(\omega)\,u(\omega)\;:\;\rho\ge0\Big\},
  \qquad u(\omega)=\big(e^{-t_1\omega},\dots,e^{-t_m\omega}\big)^T,
\end{equation}
which contains the data $Ez$ of every grid spectrum.  With
$\|x\|_\Sigma=(x^T\Sigma^{-1}x)^{1/2}$, $\chimin$ is therefore bounded from below by the
$\Sigma$-distance of $\bar C$ from $\mathcal P$.  Let $d$ be the $\Sigma$-distance of the
noise-free data $c\in\mathcal P$ from the boundary of $\mathcal P$.  Then
\begin{equation}\label{eq:pbound}
  \Prob(\chimin>0)\;\ge\;\Phi(-d),
\end{equation}
where
\begin{equation}
  \Phi(x)=\frac{1}{\sqrt{2\pi}}\int_{-\infty}^{x}e^{-s^2/2}\,ds
\end{equation}
is the standard normal cumulative distribution function.  To see this, use coordinates
in which $\Sigma$ is the unit matrix.  The ball of radius $d$ around $c$ lies in
$\mathcal P$ and touches its boundary at a point $q$.  A supporting hyperplane of the
convex set $\mathcal P$ at $q$~\cite{Rockafellar:1970} also supports the ball, so it is the
tangent plane of the ball at $q$, at distance $d$ from $c$.  $\mathcal P$ lies on one side
of this plane, and $\bar C$ falls on the other side with probability $\Phi(-d)$, in which
case $\chimin>0$.

We test Eq.~\eqref{eq:pbound} for truths with many more states than time slices
(energies in lattice units).  The spectral density
\begin{equation}\label{eq:rhov}
  \rho_V(\omega)=\omega^2\beta^3\,\frac{M^4}{(\omega^2-M^2)^2+M^2\Gamma^2},\qquad
  \beta=\sqrt{1-(0.5/\omega)^2},\qquad M=1.2,\ \Gamma=0.3,
\end{equation}
for $\omega>0.5$ has the form of a vector correlator: two-particle states above threshold
in a $P$-wave, enhanced by a resonance.  We restrict it to $\omega\le4$ and take it either
as a continuum, with weight $\rho_V(\omega_j)\Delta\omega$ on every grid node above
threshold ($\Delta\omega$ the grid spacing; 283 states), or discretized into 50
equidistant levels on $[0.5,4]$, as a proxy for a finite-volume spectrum.  The grid has 300 equidistant nodes on $[0.3,4]$, plus the 50 levels.  The data
are at $t_i=3,\dots,2+m$ for $m=4,6,8$, with relative error $r$\footnote{In lattice data the relative error
generally grows exponentially with $t$~\cite{Parisi:1984,Lepage:1989}.  We use a constant
relative error as a simpler model that illustrates the point qualitatively.} and correlation
$0.7^{|i-j|}$, i.e.\ $\Sigma=r^2\Sigma_1$ with $(\Sigma_1)_{ij}=c_ic_j\,0.7^{|i-j|}$.

\begin{figure}[t]
\includegraphics[width=0.75\textwidth]{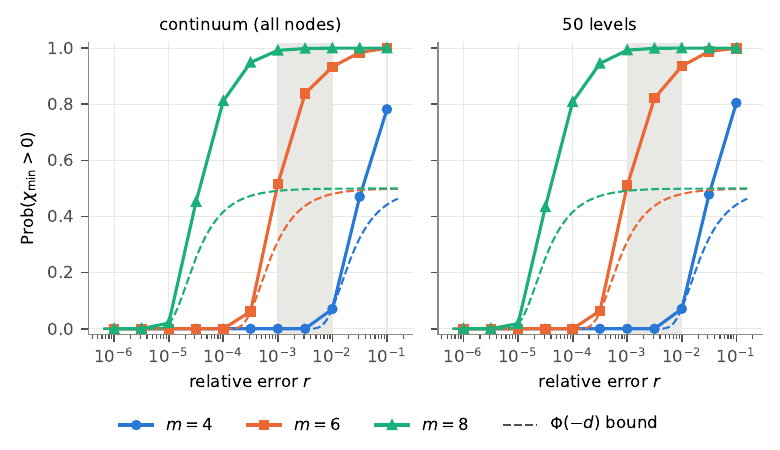}
\caption{Probability that no positive spectrum reproduces the data, $\Prob(\chimin>0)$, as a
function of the relative error $r$ of the data, for $m=4$, $6$ and $8$ time slices
(solid, Monte Carlo), and the lower bound $\Phi(-d)$ of Eq.~\eqref{eq:pbound} (dashed).
Left: continuum spectrum $\rho_V$; right: $\rho_V$ discretized into 50 levels.  Shaded:
$r=10^{-3}$--$10^{-2}$, typical of lattice correlators.}
\label{fig:chi2min}
\end{figure}

The distance $d$ sets the scale of the errors at which $\chimin>0$ sets in.
Scaling all errors by a factor $s$, $\Sigma\to s^2\Sigma$, scales $d$ to $d/s$.  $d$ is
therefore the factor by which the errors would have to be inflated for the noise-free data
to lie one standard deviation from the boundary of $\mathcal P$, much like the scale factor
$S$ by which the Particle Data Group inflates the errors of inconsistent measurements until
$\chi^2$ per degree of freedom is one~\cite{PDG:2024}.  For $N$ samples, $\Sigma\propto1/N$
and $d\propto\sqrt N$, so that $\chimin>0$ is rare for $N\gg N^*=N/d^2$ and the normal outcome for $N\ll N^*$.
For the noise model above $d=d_1/r$, where $d_1$ is the value at $r=1$.  We compute it
numerically from the characterization of the boundary of $\mathcal P$ by spectra with few
states~\cite{KarlinStudden:1966}.
Figure~\ref{fig:chi2min} shows $\Prob(\chimin>0)$ from 4000 draws of Eq.~\eqref{eq:model} at
each $r$, together with the bound \eqref{eq:pbound}.

The two panels of Fig.~\ref{fig:chi2min} agree within the Monte-Carlo errors: 283 states
or 50 levels make no difference, and the number of states is not what decides whether
$\chimin>0$.  What decides it is the precision.  For $r\ll r^*$, with $r^*=d_1$ the
relative error at which $d=1$, the data resolve that the truth lies inside $\mathcal P$
and $\chimin=0$.  For $r\gg r^*$, $\chimin>0$ is the normal outcome.  The bound
\eqref{eq:pbound} locates this crossover.  The fact that $\Prob(\chimin>0)$ rises above $\frac12$ beyond it
reflects that near its boundary $\mathcal P$ is narrower than the half-space used by the
bound.  The crossover precision is $r^*\approx1.5\times10^{-2}$,
$5\times10^{-4}$ and $2\times10^{-5}$ for $m=4$, $6$ and $8$.  $r^*$ falls by a factor
$\approx20$--$30$ per two additional time slices: additional time slices resolve
additional spectral structure only exponentially weakly, as reflected in the exponential
growth of the condition number of positive definite Hankel matrices with their
size~\cite{Beckermann:2000}.  At typical lattice precision $\chimin>0$ therefore occurs in
at least half of the cases for $m=6$ and almost always for $m=8$, as observed for resampled lattice data in
Ref.~\cite{Lawrence:2024} and for Nevanlinna--Pick interpolation in
Ref.~\cite{Fields:2025}, whereas for $m=4$ it occurs in at most 7\% of the cases here.

\subsection{The relative threshold and the statistics-dominated regime}\label{sec:stat}

We now turn from the data to the quantity of interest, a linear functional of the spectral
density,
\begin{align}
  Y=\int d\omega\,\rho(\omega)\,k(\omega),
\end{align}
on the grid $Y=k\cdot z$ with $k_j=k(\omega_j)$.  One example is $k(\omega)=e^{-t\omega}$
which gives the correlator at a time $t$.  Since $\chimin>0$ is the normal case, we bound
$Y$ by its range over the positive spectra whose $\chi^2$ lies within $\Delta\chi^2$ of the
best positive fit, a threshold relative to $\chimin^2$:
\begin{equation}\label{eq:band}
  \mathcal Z=\{z\ge0:\chi^2(z)\le\chimin^2+\Delta\chi^2\},\qquad \Delta\chi^2=1,\qquad
  \lo=\min_{z\in\mathcal Z}k\cdot z,\qquad \hi=\max_{z\in\mathcal Z}k\cdot z .
\end{equation}
$\mathcal Z$ is never empty, since it contains the best fit $z^*$.  We compute $\chimin$ and the
band exactly on the grid with the \code{jks-system} package~\cite{jks}.

\paragraph*{Positivity inactive.}
Consider an output determined by the data, $k=E^T\beta$, so that $Y=\beta\cdot Ez$ is a
linear combination of the correlator at the input times.  If the ellipsoid
$(x-\bar C)^T\Sigma^{-1}(x-\bar C)\le\Delta\chi^2$ lies inside the cone
$\mathcal P_E=\{Ez:z\ge0\}$ of data of grid spectra ($\mathcal P_E\subseteq\mathcal P$, as noted in
Sec.~\ref{sec:chi2min}), positivity is inactive: $\chimin=0$,
and the band is the range of $\beta\cdot x$ over this ellipsoid.  We call such outputs
statistics dominated.  By the Cauchy--Schwarz inequality in the metric of $\Sigma$,
$|\beta\cdot(x-\bar C)|\le(\beta^T\Sigma\beta)^{1/2}\|x-\bar C\|_\Sigma$, with equality for
$x-\bar C\propto\Sigma\beta$, so
\begin{equation}\label{eq:linear}
  \lo,\hi\;=\;\beta\cdot\bar C\;\mp\;\sqrt{\Delta\chi^2}\,\sigma_\beta,\qquad
  \sigma_\beta=(\beta^T\Sigma\beta)^{1/2}.
\end{equation}
For $\Delta\chi^2=1$ this is the standard propagated error of the data, and by
Eq.~\eqref{eq:model} the band covers the true value $\beta\cdot c$ with probability
$68.3\%$.  This is the familiar $\Delta\chi^2=1$ rule: in general the band
\eqref{eq:band} is the profile-likelihood interval
$\{y:\min_{z\ge0,\,k\cdot z=y}\chi^2(z)\le\chimin^2+\Delta\chi^2\}$ of
$Y$~\cite{JamesRoos:1975,James:2006}, since this profile is convex in $y$ with minimum
$\chimin^2$.

\paragraph*{Positivity can only tighten the band.}
In general the admissible data $x=Ez$, $z\in\mathcal Z$, lie in $\mathcal P_E$ and within
$(\chimin^2+\Delta\chi^2)^{1/2}$ of $\bar C$.  Let $p=Ez^*$ be the data of the best positive
fit, the point of $\mathcal P_E$ closest to $\bar C$, at distance $\chimin$.  For the
closed convex cone $\mathcal P_E$, $(x-p)^T\Sigma^{-1}(\bar C-p)\le0$ for all
$x\in\mathcal P_E$~\cite{Rockafellar:1970}, hence
\begin{equation}
  \|x-p\|_\Sigma^2=\|x-\bar C\|_\Sigma^2-\chimin^2+2(x-p)^T\Sigma^{-1}(\bar C-p)
  \;\le\;\Delta\chi^2 .
\end{equation}
All admissible data thus lie within $\sqrt{\Delta\chi^2}$ of $p$, and as in
Eq.~\eqref{eq:linear}
\begin{equation}\label{eq:tighten}
  [\lo,\hi]\;\subseteq\;\big[\beta\cdot p-\sqrt{\Delta\chi^2}\,\sigma_\beta,\;
  \beta\cdot p+\sqrt{\Delta\chi^2}\,\sigma_\beta\big].
\end{equation}
Positivity therefore never inflates the error of an output determined by the data
beyond Gaussian error propagation: it moves the interval to the best positive fit and can
only narrow it.\footnote{At an absolute threshold $\chi^2(z)\le F$ the same arguments give
a half-width of $\sqrt F\,\sigma_\beta$ when positivity is inactive and bound it by
$\sqrt{F-\chimin^2}\,\sigma_\beta$ otherwise, which shrinks as the fit gets worse.}  Whether
the narrower interval still covers the truth at the nominal rate is tested in
Sec.~\ref{sec:coverage}.  We use the threshold $\Delta\chi^2=1$ to quote errors
\begin{equation}\label{eq:half}
  \half=\tfrac12(\hi-\lo),
\end{equation}
which in the statistics-dominated regime are purely statistical.  There they also agree
with a jackknife analysis~\cite{Efron:1982}: if $\Sigma$ is the jackknife estimate of the
covariance of $\bar C$ from the $N$ delete-one means $C^{(b)}$, and the band is recomputed
at each $C^{(b)}$, its centre $\beta\cdot C^{(b)}$ is linear in the data and its jackknife
error is exactly $\sigma_\beta=\half$.

\subsection{Total, statistical, and systematic uncertainties}\label{sec:split}

Outside the statistics-dominated regime the band is a non-linear function of the data,
and its half-width $\half$ contains, besides the Gaussian statistical fluctuation, a
non-linear contribution caused by positivity.  We separate the two as follows.  The central
value is the centre of the band, $\hat Y=\frac12(\lo+\hi)$.  Its statistical error is
obtained by repeating the analysis on the jackknife means: at each $C^{(b)}$ the band
\eqref{eq:band} is recomputed with the same $\Sigma$ and its own $\chimin$, giving
$\hat Y^{(b)}$, and
\begin{equation}\label{eq:sstat}
  \sstat^2=\frac{N-1}{N}\sum_{b=1}^N\big(\hat Y^{(b)}-\langle\hat Y\rangle\big)^2,\qquad
  \langle\hat Y\rangle=\frac1N\sum_{b=1}^N\hat Y^{(b)} .
\end{equation}
The total and systematic errors are
\begin{equation}\label{eq:sys}
  \stot=\max(\sstat,\half),\qquad \ssys^2=\stot^2-\sstat^2 ,
\end{equation}
so that $\stot^2=\sstat^2+\ssys^2$.  The maximum covers the case $\sstat>\half$, in which
the centre moves between resamples by more than the band is wide.

The rationale is the following.  The half-width $\half$ already contains the statistical
error (Sec.~\ref{sec:stat}), so $\half$ and $\sstat$ must not be added in quadrature as doing so would
overstate the error of an output determined by the data by a factor $\sqrt2$.  Instead, the
jackknife identifies the part of $\half$ that responds linearly to the data, which is
therefore Gaussian to the extent that the data are and carries their correlations with
other quantities.  $\ssys$ is the remainder: the non-Gaussian part of the error.  In the
statistics-dominated regime $\sstat=\half$ and $\ssys=0$ (Sec.~\ref{sec:stat}), and
$\stot$ is the Gaussian propagated error.

\begin{figure}[t]
\includegraphics[width=\textwidth]{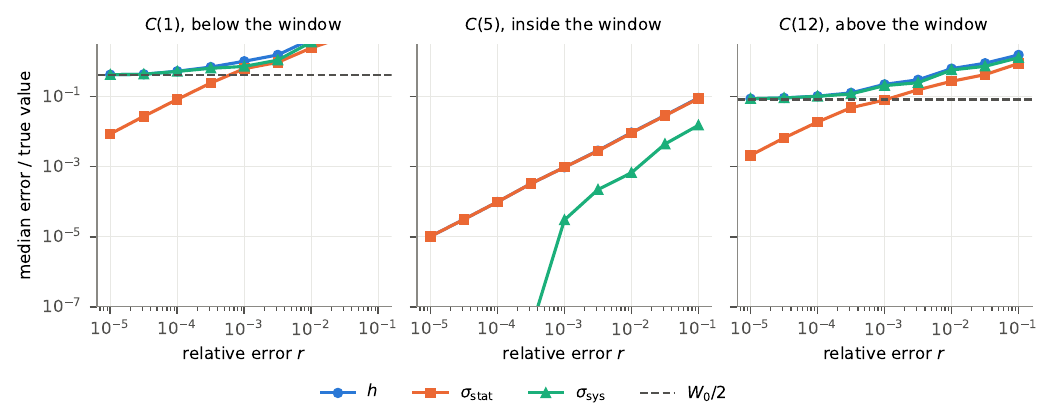}
\caption{Half-width $\half$, statistical error $\sstat$ and systematic error $\ssys$,
relative to the true value, for the continuum spectrum $\rho_V$ with data at
$t_i=3,\dots,8$, as a function of the relative error $r$.  The medians are computed over 20 ensembles of
$N=50$ samples with $\Sigma$ the jackknife covariance.  The dashed line shows the half-width $W_0/2$ of the
exact-data band.  The left and right panels show the outputs $C(1)$ and $C(12)$ outside the input window,
the middle panel shows $C(5)$ inside it, where $\ssys=0$ for $r<r^*$.}
\label{fig:split}
\end{figure}

As $N\to\infty$, $\Sigma\to0$, $\sstat\propto N^{-1/2}\to0$, and the band tends to the range
$[\lo_0,\hi_0]$ of $Y$ over the positive spectra that reproduce the noise-free data,
$\{z\ge0:Ez=c\}$.  Hence $\ssys\to W_0/2$ with $W_0=\hi_0-\lo_0$, the ambiguity that
positivity leaves for exact data.  For $W_0>0$, $\ssys$ is then a systematic error in the
usual sense, one that does not decrease with the statistics.  Figure~\ref{fig:split} shows this for
the continuum truth of Fig.~\ref{fig:chi2min} at $m=6$ ($t_i=3,\dots,8$): inside the input
window $\ssys=0$ for $r<r^*$, while outside it $\sstat\propto r$ and $\ssys$ approaches
$W_0/2$.  For a truth on the boundary of $\mathcal P$, in contrast, the representing
spectrum is unique~\cite{KarlinStudden:1966}, $W_0=0$, and $\ssys$ decreases with the
statistics like $\sstat$.  For a two-state truth at $m=6$ (energies $0.5$ and $1.0$,
weights $1$ and $0.5$) our numerical experiments show $\ssys/\half\approx0.5$--$0.8$ for outputs outside the window
at all $r$ from $10^{-5}$ to $10^{-1}$.

A physical
spectrum has infinitely many states, so its noise-free data lie in the interior of
$\mathcal P$ and $W_0>0$.  As shown in Sec.~\ref{sec:chi2min}, however, the data resolve
this only when their relative error is below the crossover precision $r^*$.  For
$r\gg r^*$ the noise-free data are indistinguishable, within the errors, from those of a
spectrum with fewer than $m/2$ states, which lie on the boundary of $\mathcal P$.  In
this regime $\ssys$ behaves as for a boundary truth and decreases with the statistics: it
is the non-Gaussian part of the statistical error, caused by positivity.  Only for
$r\ll r^*$, when the data resolve that the truth lies in the interior, does $\ssys$
approach $W_0/2$ and become a systematic error that further statistics does not reduce.

\subsection{Coverage}\label{sec:coverage}

We test whether $\hat Y\pm\stot$ covers the true value of $Y$ with at least the nominal
probability.  Each test repeats the complete analysis of Secs.~\ref{sec:stat}
and~\ref{sec:split} on 1000 independent ensembles of $N=50$ Gaussian samples each, drawn with covariance
$N\Sigma$ so that their mean has the covariance $\Sigma$ of the noise model of
Sec.~\ref{sec:chi2min}, for data at $t_i=3,\dots,8$ with relative errors $r=10^{-4}$,
$10^{-3}$ and $10^{-2}$ and on the grid of Sec.~\ref{sec:chi2min}.  In the analysis $\Sigma$
is replaced by its jackknife estimate.  We use two truths: the continuum $\rho_V$ of Fig.~\ref{fig:chi2min}, whose
data lie in the interior of $\mathcal P$, and a two-state spectrum (energies $0.5$ and $1.0$,
weights $1$ and $0.5$), whose data lie on its boundary.  The outputs are the correlator
$C(t)$ at $t=3$, $5$ and $8$ inside the input window and at $t=1$ and $12$ outside it, as
well as smeared spectral densities and smeared step functions,
\begin{equation}\label{eq:kernels}
  k(\omega)=\delta_\tau(\omega-\omega_c)=\frac{e^{-(\omega-\omega_c)^2/2\tau^2}}{\sqrt{2\pi}\,\tau},\qquad
  k(\omega)=\theta_\tau(\omega_c-\omega)=\tfrac12\Big[1+\tanh\frac{\omega_c-\omega}{\tau}\Big],
\end{equation}
the latter giving the spectral weight below $\omega_c$, at $(\omega_c,\tau)=(1.2,0.1)$, $(1.2,0.3)$ and
$(2.0,0.3)$ for $\delta_\tau$ and $(0.8,0.1)$ and $(0.8,0.3)$ for $\theta_\tau$.  For an output
determined by the data and with positivity inactive, $\stot=\sigma_\beta$, and with $\Sigma$
estimated from $N$ samples the interval $\pm\sigma_\beta$ covers with the probability of
Student's $t$ distribution with $N-1$ degrees of freedom~\cite{Student:1908}, $0.678$ for
$N=50$, which is the reference value for these outputs.

\begin{figure}[t]
\includegraphics[width=\textwidth]{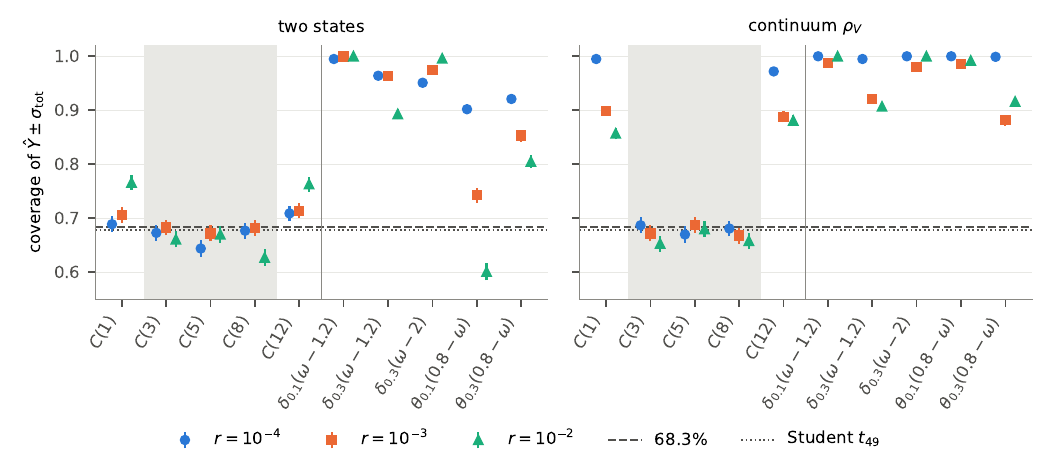}
\caption{Coverage of $\hat Y\pm\stot$ for the two-state truth (left) and the continuum
$\rho_V$ (right), for data at $t_i=3,\dots,8$ with relative error $r$; 1000 ensembles of
$N=50$ samples per point.  Shaded: outputs inside
the input window.  Dashed: $68.3\%$; dotted: Student's $t$ with 49 degrees of freedom.}
\label{fig:coverage}
\end{figure}

Figure~\ref{fig:coverage} shows the result.  Inside the input window the coverage scatters
around the Student reference, $0.627$--$0.688$.  The lowest values arise from the estimate of
$\Sigma$ from $N=50$ samples, which enters the band non-linearly when positivity is active:
for the two-state truth the same outputs cover $0.69$--$0.71$ with the exact covariance and
$0.68$--$0.69$ with $N=200$ samples ($10^4$ and $5\times10^3$ ensembles).  Outside the window, and for the smeared kernels, the systematic part of the
error dominates and the coverage is mostly well above nominal, up to $1.00$, since the truth
can lie anywhere in a band whose width is set by $W_0$ rather than by the statistics. The
exception is the sharp step $\theta_{0.1}(0.8-\omega)$ of the two-state truth, for which the
truth lies at the lower end of the band and the coverage falls from $0.90$ at $r=10^{-4}$ to
$0.60\pm0.015$ at $r=10^{-2}$.  In the limit of vanishing statistical error the band contains
the truth by construction, and the under-coverage at large $r$ is an effect of the statistical
error on the estimate of the error, which disappears as the statistics is increased.  Outside the window $\pm\sstat$ alone covers in as few as $0\%$ of the cases, so the
systematic part is indispensable.

\section{Prior information on the spectral density}\label{sec:prior}

Information on the spectral density beyond positivity enters as additional convex
constraints.  We consider bounds on the size of $\rho$ (Sec.~\ref{sec:densitybounds}) and on its
support (Sec.~\ref{sec:support}). The latter connects the method to the bounding method used
for the long-distance correlator.

\subsection{Two-sided bounds on the spectral density}\label{sec:densitybounds}

Positivity can be supplemented by bounds $\rho_{\min}(\omega)\le\rho(\omega)\le\rho_{\max}(\omega)$, for
example from perturbation theory, which predicts $\rho$ above some energy $E_{\rm pt}$ up to an
uncertainty, or from a loose upper bound, which limits the weight at energies that the data
barely constrain.  On the grid the bounds
apply to the weights, $l_j\le z_j\le u_j$, with $l_j$ and $u_j$ the integrals of $\rho_{\min}$ and
$\rho_{\max}$ over the bin of node $j$.  In a finite volume, where $\rho$ is a sum of
$\delta$-functions, such bounds are meaningful only for bins wider than the level spacing.

The construction of Sec.~\ref{sec:interp} carries over with the cone $\mathcal P_E$ replaced by the
image $\{Ez:l\le z\le u\}$ of the bounds, which is again closed and convex. Only the
characterization of its boundary by spectra with few states is specific to positivity.  $\chimin$ becomes a
bounded-variable least-squares problem~\cite{StarkParker}, the bound \eqref{eq:pbound} and the
argument leading to Eq.~\eqref{eq:tighten} use only convexity, and the split of
Sec.~\ref{sec:split} applies with $W_0$ the width of the exact-data band under the bounds.

\begin{table}[t]
\caption{Half-width $W_0/2$ of the exact-data band relative to the true value, for the continuum
spectrum $\rho_V$ with data at $t_i=3,\dots,8$ and four choices of bounds.}
\label{tab:bounds}
\begin{tabular}{l c c c c c}
\toprule
bounds & $C(1)$ & $C(12)$ & $\delta_{0.3}(\omega-2)$ & $\delta_{0.3}(\omega-1.2)$ & $\theta_{0.1}(0.8-\omega)$\\
\midrule
$\rho\ge0$ only                         & 0.41  & 0.085 & 0.67 & 0.047 & 0.45\\
$z_j\le3\max_jz_j^{\rm true}$           & 0.20  & 0.081 & 0.65 & 0.045 & 0.45\\
$(1\mp5\%)\rho_V$ for $\omega\ge2$      & 0.008 & 0.051 & 0.34 & 0.018 & 0.40\\
$(1\mp20\%)\rho_V$ for $\omega\ge2$     & 0.010 & 0.052 & 0.35 & 0.020 & 0.41\\
\bottomrule
\end{tabular}
\end{table}

Table~\ref{tab:bounds} shows the effect on $W_0$ for the continuum truth of
Fig.~\ref{fig:chi2min} with data at $t_i=3,\dots,8$.  We compare positivity alone with a loose
cap, three times the largest true weight on every node, and with bounds $(1\mp\epsilon)\rho_V$ for
$\omega\ge E_{\rm pt}=2$, which stand in for a perturbative prediction with relative uncertainty
$\epsilon$.  Bounds at high energies reduce the systematic error of outputs that are sensitive
to them by large factors, by a factor 40--50 for $C(1)$, while outputs determined by the spectrum at
low energies change little, and inside the input window $\ssys$ remains negligible.  The width
$\epsilon$ hardly matters: once the spectrum above $E_{\rm pt}$ is constrained, the remaining
ambiguity comes from energies below $E_{\rm pt}$.  At $r=10^{-5}$ the measured $\ssys$ agrees with
$W_0/2$ to within $7\%$ for all four choices. At $r=10^{-3}$ the reduction is smaller ($\ssys/C(1)$ from $0.84$ to
$0.04$), since there the non-Gaussian statistical part of the error dominates
(Sec.~\ref{sec:split}).  The smaller errors do not cost coverage: repeating the test of
Sec.~\ref{sec:coverage} with the $\epsilon=20\%$ bounds on 400 ensembles per $r$ gives
$0.83$--$1.00$ for all outputs outside the input window (positivity alone: $0.84$--$1.00$ on the
same ensembles), and inside the window the coverage is unchanged.

\subsection{Support of the spectral density and the bounding method}\label{sec:support}

The lower edge $\omega_1$ of the support controls the prediction of the correlator at times
beyond the inputs.  If $\omega_1$ lies below the lowest energy $E_0$ of the states, positive
spectra may carry weight that decays as slowly as $e^{-\omega_1t}$. For $\omega_1=0$ positivity
alone only implies that $C(t)$ does not increase.  If $E_0$ is known, for example the lowest finite-volume
two-particle energy~\cite{Luscher:1991}, the support can be restricted to $\omega_1=E_0$.  Then
$C(t)e^{E_0t}=\int d\omega\,\rho(\omega)e^{-(\omega-E_0)t}$ does not increase either, so that
$C(t)\le C(T)e^{-E_0(t-T)}$ for $t\ge T$.  Positivity also implies that $\log C(t)$ is convex,
since $C(\frac{t+s}2)^2\le C(t)C(s)$ by the Cauchy--Schwarz inequality, so that
$C(t)\ge C(T)e^{-E^*_T(t-T)}$ for integer $t\ge T$ with the effective mass
$E^*_T=\log(C(T)/C(T+1))$.  These two inequalities are the upper and lower bounds of the
bounding method~\cite{Lehner:2016talk,RBC:2018,BMW:2017slope}, which replaces $C(t)$ for $t\ge T$ by
$C(T)e^{-E_0(t-T)}$ and $C(T)e^{-E^*_T(t-T)}$, respectively, and quotes their average once the
two agree within errors.

The method of Sec.~\ref{sec:interp} with support $[E_0,\infty)$ uses the same information.
With only the inputs $C(T)$ and $C(T+1)$, the band of $C(t)$, $t\ge T+1$, for exact data is
\begin{equation}\label{eq:twoband}
  \big[\lo,\hi\big]=\big[\,C(T)\,e^{-E^*_T(t-T)},\;C(T+1)\,e^{-E_0(t-T-1)}\,\big],
\end{equation}
the lower bound of the bounding method at $T$ and its upper bound at $T+1$.  Both ends are
bounds by the two inequalities above.  The lower end is attained by the single exponential with
energy $E^*_T\ge E_0$. The upper end is approached by weight at $E_0$ that reproduces $C(T+1)$,
plus weight at an energy $\Omega\to\infty$ that makes up the rest of $C(T)$ and does not
contribute at later times.  Inputs at earlier times can only narrow the band.

With noisy data and positivity inactive, the band ends are the extremes of the two functions in
Eq.~\eqref{eq:twoband} over the $\Delta\chi^2=1$ ellipse of the inputs, i.e.\ the two bounds
shifted outwards by their linearly propagated errors $\sigma_{\rm lo}$ and $\sigma_{\rm up}$.  The
half-width is then
\begin{equation}\label{eq:hbound}
  \half=\tfrac12(C_{\rm up}-C_{\rm lo})+\tfrac12(\sigma_{\rm lo}+\sigma_{\rm up}),
\end{equation}
with $C_{\rm lo}$ and $C_{\rm up}$ the two ends of Eq.~\eqref{eq:twoband}.  Since the two bounds are
strongly correlated, the error of their average is close to $\frac12(\sigma_{\rm lo}+\sigma_{\rm up})$,
so that the error of the method of Sec.~\ref{sec:interp} equals that of the bounding method up
to half the remaining gap between the bounds, which it keeps as a systematic error.  For the
50-level spectrum of Fig.~\ref{fig:chi2min}, with $t=T+6$ and the noise model of
Sec.~\ref{sec:chi2min}, we find $\half/\sigma_{\rm avg}=1.10$ when half the gap is $0.17\,\sigma_{\rm avg}$
($T=40$, $r=0.1$) and $1.75$ when it is $0.81\,\sigma_{\rm avg}$ ($T=26$, $r=0.1$), where
$\sigma_{\rm avg}$ is the error of the average of the bounds.

\section{Conclusions}\label{sec:conclusions}

Bounds on linear functionals of the spectral density from positivity and a $\chi^2$
measure of compatibility with the data can be quoted as a central value with statistical and
systematic errors that have a definite meaning.  The starting point is that, at typical
precision, no positive spectrum reproduces noisy data exactly.  The probability of
$\chimin>0$ is bounded from below by the $\Sigma$-distance $d$ of the noise-free data from the
boundary of the set of data of positive spectra, and $\chimin>0$ is the normal outcome when
$d\ll1$, i.e.\ for $N\ll N^*=N/d^2$ samples.  The threshold on $\chi^2$ is therefore set relative
to the best positive fit.  With $\Delta\chi^2=1$ the resulting interval is the
profile-likelihood interval of the output.  It reproduces Gaussian error propagation when
positivity does not act, and positivity can only narrow it.

The statistical error is the jackknife error of the centre of the interval, computed with the
threshold relative to the best fit at every resample.  It is Gaussian to the extent that the
data are and carries their correlations with other quantities.  The systematic error is the
remaining, non-Gaussian part of the half-width.  It vanishes where positivity does not act,
and it becomes a systematic error that does not decrease with the statistics only when the
data are precise enough to resolve that the true spectrum is not a spectrum with few states.
Otherwise it decreases with the statistics like the statistical error.  In coverage tests with
correlators, smeared spectral densities and smeared step functions, the total error covers the
truth at the nominal rate for outputs determined by the data and mostly conservatively for the
others.  The exceptions are effects of the statistical error on the estimate of the error,
from the estimate of the covariance or from a truth at the edge of the interval, which
disappear as the statistics is increased.

Information beyond positivity enters as further convex constraints, to which the construction
carries over unchanged.  Bounds on the spectral density at high energies, such as a
perturbative prediction, reduce the systematic error of outputs sensitive to these energies by
large factors without loss of coverage.  A known lowest energy restricts the support, and with
it the method uses the same information as the bounding method for the long-distance
correlator. Its error agrees with that of the bounding method up to half the remaining gap
between the bounds, which it keeps as a systematic error.  The prescription is implemented in
the \code{jks-system} package~\cite{jks}.  Natural next steps are its application to lattice data,
in particular to the long-distance part of the hadronic vacuum polarization, and a treatment of
the correlations of the systematic error between different outputs, which matter for sums of
outputs such as time-momentum integrals.

\begin{acknowledgments}
  We are grateful to our colleagues of the RBC/UKQCD collaborations,
  in particular Ryan Abbott, Norman Christ, Sarah Fields, and Patrick Oare as well
  as Will Jay and Matteo Saccardi for useful discussions.  We
  acknowledge the use of Anthropic's Claude Opus 5.5 for help with the
  contents and phrasing of this manuscript.  CL led the research and
  all results and statements in this manuscript have been
  independently verified by CL.
\end{acknowledgments}

\bibliography{refs}

\end{document}